\documentclass[aps,prd,twocolumn,showpacs,superscriptaddress,groupedaddress]{revtex4-1}  
\usepackage{amssymb}
\usepackage{amsfonts}
\usepackage{amsmath,array}
\usepackage{epsfig}
\usepackage{graphicx}
\usepackage{epstopdf} 
\usepackage{dcolumn}
\usepackage{bm}
\usepackage{lineno}
\usepackage{multirow}
\usepackage{color}
\usepackage{bm}
\usepackage[utf8]{inputenc}
\usepackage[perpage,symbol]{footmisc}
\usepackage{subfigure}
\setfnsymbol{wiley}

\newcommand{\jpsi}{J/\psi}
\newcommand{\jpsito}{J/\psi\rightarrow}

\newcommand{\llb}{\Lambda\bar{\Lambda}}

\newcommand{\pbar}{\bar{p}}
\newcommand{\nbar}{\bar{n}}
\newcommand{\pip}{\pi^{+}}
\newcommand{\pim}{\pi^{-}}
\newcommand{\piz}{\pi^{0}}
\newcommand{\lam}{\Lambda}
\newcommand{\lbar}{\bar{\Lambda}}

\newcommand{\gam}{\gamma}
\newcommand{\ra}{\rightarrow}

\newcommand{\ajpsi}{\alpha_{J/\psi}}
\newcommand{\afp}{\alpha_{+}}
\newcommand{\afm}{\alpha_{-}}
\newcommand{\afz}{\bar{\alpha}_{0}}
\newcommand{\meth}{Appendix \ref{sec:meth}}
\newcommand{\Njpsi}{(1310.6\pm7.0)\times10^6} 

\begin{document}

\title{Polarization and Entanglement in Baryon-Antibaryon Pair Production in
Electron-Positron Annihilation}

\author{
\begin{small}
\begin{center}
M.~Ablikim$^{1}$, M.~N.~Achasov$^{10,d}$, S. ~Ahmed$^{15}$, M.~Albrecht$^{4}$, M.~Alekseev$^{55A,55C}$, A.~Amoroso$^{55A,55C}$, F.~F.~An$^{1}$, Q.~An$^{52,42}$, Y.~Bai$^{41}$, O.~Bakina$^{27}$
% ...
\\
\vspace{0.2cm}
(BESIII Collaboration)\\
\vspace{0.2cm} {\it
$^{1}$ Institute of High Energy Physics, Beijing 100049, People's Republic of China\\
% ...
$^{4}$ Bochum Ruhr-University, D-44780 Bochum, Germany\\
% ...
$^{10}$ G.I. Budker Institute of Nuclear Physics SB RAS (BINP), Novosibirsk 630090, Russia\\
% ...
$^{15}$ Helmholtz Institute Mainz, Johann-Joachim-Becher-Weg 45, D-55099 Mainz, Germany\\
% ...
$^{27}$ Joint Institute for Nuclear Research, 141980 Dubna, Moscow region, Russia\\
% ...
$^{41}$ Southeast University, Nanjing 211100, People's Republic of China\\
$^{42}$ State Key Laboratory of Particle Detection and Electronics, Beijing 100049, Hefei 230026, People's Republic of China\\
% ...
$^{52}$ University of Science and Technology of China, Hefei 230026, People's Republic of China\\
% ...
$^{55}$ (A)University of Turin, I-10125, Turin, Italy; (B)University of Eastern Piedmont, I-15121, Alessandria, Italy; (C)INFN, I-10125, Turin, Italy\\
% ...
$^{d}$ Also at the Novosibirsk State University, Novosibirsk, 630090, Russia\\
% ...
}
\end{center}
\vspace{0.4cm}
\end{small}
}
\noaffiliation{}
\begin{abstract}
Using a sample of $1.31\times 10^9$ $\jpsi$ events collected with the BESIII detector, we report the first observation of spin polarization of $\lam$ and $\lbar$ hyperons from the coherent production in the $\jpsito\llb$ decay.  We measure the phase between the 
hadronic form factors to be $\Delta\Phi=(42.4\pm0.6\pm0.5)^\circ$.
The decay parameters 
 for $\lam\ra p\pim$ ($\afm$), $\lbar\ra\pbar\pip$ ($\afp$)
and $\lbar\ra\nbar\piz$ ($\afz$)
are measured to be $\afm=0.750\pm0.009\pm0.004$, $\afp=-0.758\pm0.010\pm0.007$ and $\afz=-0.692\pm0.016\pm0.006$, respectively.
The obtained value of $\afm$  is higher by $(17\pm 3)\%$ than the current world average.  In addition,  the $CP$ asymmetry $A_{CP}=(\afm+\afp)/(\afm-\afp)$
of $-0.006\pm0.012\pm0.007$ is extracted with substantially improved precision.
The ratio 
$\bar{\alpha}_{0}/\alpha_{+} = 0.913\pm 0.028  \pm 0.012$ is also measured.
\end{abstract}
\date{\today}
\maketitle
\section{Introduction}
The well-defined and simple 
initial state makes baryon-antibaryon pair
production
 at an electron-positron collider an ideal system to test fundamental
symmetries in the baryon sector, in particular when the probability of
the process is enhanced
by a resonance such as the $J/\psi$~\cite{Kopke:1988cs}.
The spin orientations of the baryon and
    antibaryon are entangled and, for spin one-half baryons, the pair
    is produced either with the same or opposite helicities. The 
    transition amplitudes to the respective spin states
    can acquire a relative phase due to the strong interaction in the
    final state, leading to a time-reversal-odd observable:
    a transverse spin polarization of the
    baryons~\cite{Cabibbo:1961sz,Brodsky:2003gs}. 
     This effect has previously been
    neglected for the baryon pairs from $\jpsi$
    decays~\cite{Faldt:2017kgy}, and there is no prediction for this
    polarization.

    Here, we observe baryon polarization for the first
    time in an electron-positron reaction.  The baryon studied is the
    $\Lambda$ hyperon,
which decays via a parity
    non-conserving weak process. 
    The Lambda polarisation is extracted from the angular distribution of
    its decay products; from this, the production phase is determined to be
    $(42.4\pm0.6\pm0.5)^\circ$.  The observed polarization allows the
    simultaneous determination of the $\Lambda$ and $\bar\Lambda$
    decay asymmetries from the events, in which all decay products are
    measured.  A sensitive $CP$ symmetry test in the strange baryon
    sector is performed by directly comparing the asymmetry parameters
    for $\Lambda\to p\pi^-$ and $\bar\Lambda\to \bar p\pi^+$.

    Of
    major importance is our result for the $\Lambda\to p\pi^-$
    asymmetry parameter.
 Nearly all
    experiments making use of the $\Lambda$ polarization (some
    examples are given in
    Refs.~\cite{Bunce:1976yb,Pondrom:1985aw,Bonner:1988au,Luk:1988as,Barnes:1996si,Chakravorty:2003je,ATLAS:2014ona,STAR:2017ckg})
    use this decay for reconstruction and the $\Lambda\to p\pi^-$
    asymmetry parameter in the determination of the polarization. To
    determine the polarization from the product of the polarization
    and the asymmetry parameter, all these studies assume an asymmetry
    parameter of $0.642\pm0.013$, the world-average value
    established from results spanning
    1963--75~\cite{Cronin:1963zb,Overseth:1967zz,Dauber:1969hg,Cleland:1972fa,Astbury:1975hn}.

\begin{figure}
\centering
\includegraphics[width=0.49\textwidth]{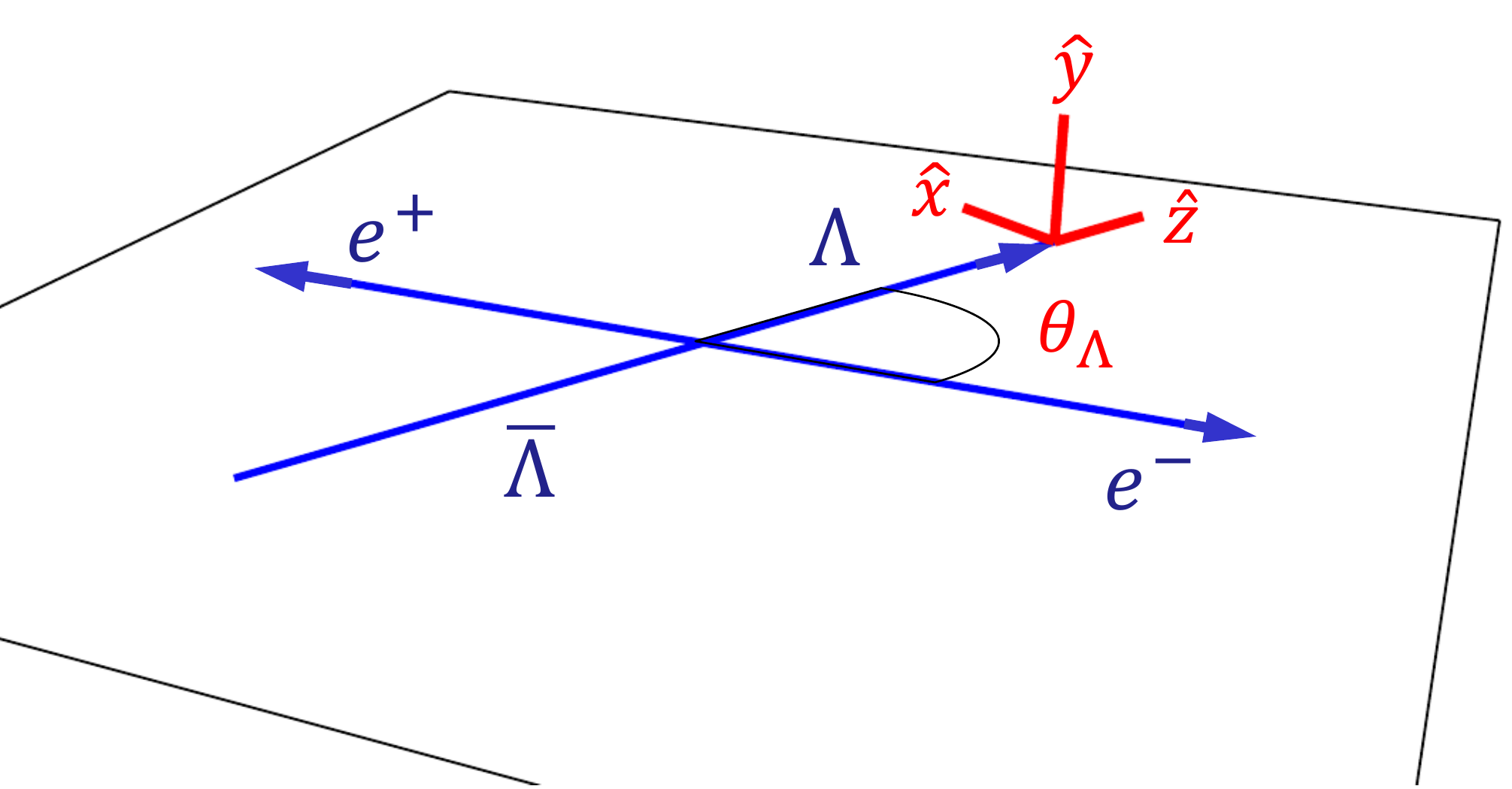}
\caption[]{Kinematics of the reaction $e^+e^-\to\jpsito\llb$ in the overall
center-of-mass system.
The $\Lambda$ particle is emitted in
the $\hat{z}$ direction at an angle $\theta_\Lambda$ with respect to
the $e^-$ direction, and the $\bar\Lambda$ is emitted in the opposite
direction.  The hyperons are polarized in the direction perpendicular
to the reaction plane ($\hat{y}$).  The hyperons are reconstructed,
and the polarization is determined by measuring their decay products:
(anti-)nucleons and pions.
\label{fig:exp}}
\end{figure}

The principle of the measurement is illustrated in
Fig.~\ref{fig:exp}. In the collision of an electron and positron, a
$\jpsi$ resonance is produced at rest in a single photon annihilation
process, and it subsequently decays into a $\Lambda\bar\Lambda$
pair: $e^+e^-\to\jpsito\llb$.
The  transition between the initial electron-positron pair
and the final baryon-antibaryon pair includes helicity conserving and
-flip amplitudes~\cite{Dubnickova:1992ii,Gakh:2005hh,Czyz:2007wi,Faldt:2013gka,Faldt:2016qee}. Since
the electron mass is negligible in comparison to the $\jpsi$ mass, the
initial electron and positron helicities have to be opposite. This
implies that the angular distribution and polarization of the produced
$\Lambda$ and $\bar\Lambda$ particles can be described uniquely by only two
real parameters: $\alpha_\psi$ --- the $\jpsito\llb$ angular distribution parameter,
and $\Delta\Phi$ --- the phase between the two helicity amplitudes.
The value of the parameter $\alpha_\psi$ is well
known~\cite{Bai:1998fu,Ablikim:2012qn,Ablikim:2017tys}, but the phase
$\Delta\Phi$ has never been considered before.  If $\Delta\Phi\ne0$, the
$\Lambda$ and $\bar\Lambda$ will be polarized in the direction
perpendicular to the production plane. The magnitude of the
polarization depends on the angle ($\theta_\Lambda$) between the
$\Lambda$ and the electron beam direction in the $\jpsi$ rest frame
(see Fig.~\ref{fig:exp}).

The polarization of
the weakly decaying particles, such as the $\Lambda$ {hyperons}, can be determined using 
the angular distribution of the daughter particles.
{For example,  for the $\Lambda\to p\pi^-$ decay, the $\Lambda$ hyperon
polarization vector, ${\bf P}_\Lambda$, is given by the angular distribution of the
daughter protons via $\frac{1}{4\pi}\left(1+\alpha_- {\bf P}_\Lambda\cdot\hat{\bf
  n}\right)$}, where $\hat{\bf n}$ is the unit vector along the proton momentum
in the $\Lambda$ rest frame and $\alpha_-$ is the
asymmetry parameter of the decay~\cite{Lee:1957qs}. The
corresponding parameters
 $\afp$ for $\lbar\ra\pbar\pip$, $\alpha_0$ for $\Lambda\ra
n\piz$, and $\afz$ for $\lbar\ra\nbar\piz$ are defined in the same way~\cite{pdg16}. 
The joint angular distribution of $\jpsi\to \llb$ ($\Lambda \to f$ and $\bar\Lambda \to \bar{f}$, $f= p \pi^-$ or $n \pi^0$)  depends on the $\Lambda$
and $\bar\Lambda$
polarization and spin correlation  of the $\llb$ pair via the  
$\alpha_\psi$ and  $\Delta\Phi$ parameters.
In particular, the joint angular distribution of the decay chain $\jpsi\to  
(\Lambda \to  p \pi^-)(\bar\Lambda \to  \bar p \pi^+) $
 can be expressed as~\cite{Faldt:2017kgy}:
\begin{widetext}
\begin{equation}
\begin{split}
{\cal{W}}({\boldsymbol{\xi}};\alpha_\psi,\Delta\Phi,\alpha_-,\alpha_+)=&1+{{{\alpha_\psi}}}{\cos^2\!\theta_\Lambda}\\
+&\ {{\alpha_-\alpha_+}}
\left[
{\sin^2\!\theta_\Lambda}\left(n_{1,x}n_{2,x} -{{\alpha_\psi}} n_{1,y}n_{2,y}\right)
+\left({\cos^2\!\theta_\Lambda}+{{\alpha_\psi}}\right)
n_{1,z}n_{2,z}\right]\\
+&\ {{\alpha_-\alpha_+}}\sqrt{1-{{\alpha_\psi}}^2}\cos({{\Delta\Phi}})
 {\sin\theta_\Lambda\cos\theta_\Lambda}\left(n_{1,x}n_{2,z}+n_{1,z}n_{2,x}\right)\\
+&\sqrt{1-{{\alpha_\psi}}^2}\sin({{\Delta\Phi}})\ {\sin\theta_\Lambda\cos\theta_\Lambda}
\left({{\alpha_-}}n_{1,y}
+{{\alpha_+}}n_{2,y}\right),\label{jointangular}
\end{split}
\end{equation}
\end{widetext}
where $\hat{\bf n}_1$ ($\hat{\bf n}_2$) is the unit vector in the
direction of the nucleon (antinucleon) in the rest
frame of $\lam$ ($\lbar$).  The components of these
vectors are expressed using a common $(\hat{x},\hat{y},\hat{z})$
coordinate system with the orientation shown in
Fig.~\ref{fig:exp}. The $\hat{z}$ axis in the $\lam$ and $\lbar$ rest
frames is oriented along the $\lam$ momentum ${\bf p}_\Lambda$ in the
$\jpsi$ rest system. The $\hat{y}$ axis is perpendicular to the
reaction plane and oriented along the vector ${\bf k}_{-}\!\times{\bf
p}_\Lambda$, where ${\bf k}_{-}$ is the electron beam momentum
in the $\jpsi$ rest system.
The variable ${\boldsymbol{\xi}}$ denotes the tuple $(\theta_\Lambda,\hat{\bf
n}_1,\hat{\bf n}_2)$, a set of kinematic variables which uniquely
specify an event configuration.  The terms multiplied by
${{\alpha_-\alpha_+}}$ in Eq.~\eqref{jointangular} represent the
contribution from
$\llb$ spin correlations, while the terms multiplied by $\alpha_-$ and
$\alpha_+$ separately represent the contribution from the polarization.  
The presence of
all three contributions in Eq.~\eqref{jointangular} enables an unambiguous
determination of the parameters $\alpha_\psi$ and $\Delta\Phi$ and the
decay asymmetries $\alpha_-$, $\alpha_+$.  
If  $\bar\Lambda$ is reconstructed via its $\bar
n \pi^0$ decay, the parameters 
$\alpha_\psi$, $\Delta\Phi$ and the
decay asymmetries $\alpha_-$ and $\bar\alpha_0$ can be determined
independently,
since the corresponding
angular distribution  is obtained by replacing $\alpha_+$ by $\bar\alpha_0$ 
and interpreting $\hat{\bf n}_2$ as the antineutron direction in 
Eq.~\eqref{jointangular}.

\section{Analysis}
The analysis is based on $\Njpsi$ $\jpsi$
events~\cite{Ablikim:2016fal} collected with the BESIII detector,
which
is described in detail in
Ref.~\cite{Ablikim:2009aa}.
The $\lam$ hyperons are reconstructed using their 
$p \pim$ decays and the $\lbar$ hyperons 
using their $\pbar\pip$ or $\nbar\piz$ decays.  The event
reconstruction and selection procedure are described in \meth.
 The resulting data samples are essentially background free,
as shown in Figs.~\ref{LLbarcut} and \ref{NbarPiAlgcut}.  A
Monte Carlo (MC) simulation including all known $\jpsi$ decays
 is used to determine the background
contribution.  The sizes of the final data samples are 420,593 and
47,009 events with an estimated background of $399\pm20$ and
$66.0\pm8.2$ events for the $p\pi^-\bar{p}\pi^+$ and
$p\pi^-\bar{n}\pi^0$ final states, respectively. For each event the
full set of the kinematic variables ${\boldsymbol{\xi}}$ is
reconstructed.

 The free parameters describing the angular distributions  
for the two data sets ---
$\alpha_{\psi}$, $\Delta\Phi$, $\alpha_{-}$, $\alpha_{+}$,
and $\bar\alpha_{0}$ ---
 are determined from a simultaneous unbinned maximum likelihood fit. In the fit, the likelihood function is constructed
from the probability density function ${\cal P}(\boldsymbol{\xi}^{(i)})={\cal
  C}(\alpha_\psi,\Delta\Phi,\alpha_-,\alpha_2)\times$
${\cal
  W}({\boldsymbol{\xi}^{(i)}};\alpha_\psi,\Delta\Phi,\alpha_-,\alpha_2)$ with $\alpha_2=\alpha_+$ 
and $\alpha_2=\bar\alpha_0$ for the $p\pi^-\bar{p}\pi^+$ and
$p\pi^-\bar{n}\pi^0$ data sets, respectively. The final configuration
of an event $i$ is characterized by the vector
$\boldsymbol{\xi}^{(i)}$, and  ${\cal
W}({\boldsymbol{\xi}^{(i)}};\alpha_\psi,\Delta\Phi,\alpha_-,\alpha_2)$
is given by Eq.~\eqref{jointangular}.
The normalization of the probability function, ${\cal
C}(\alpha_\psi,\Delta\Phi,\alpha_-,\alpha_2)$, is determined for each
parameter set using a sum of the weights ${\cal
W}({\boldsymbol{\xi}^{(m)}};\alpha_\psi,\Delta\Phi,\alpha_-,\alpha_2)$
for an ensemble ${\boldsymbol{\xi}^{(m)}}$ of isotropically
generated MC events (with ${\cal W}({\boldsymbol{\xi};0,0,0,0})\equiv1$). 
The generated 
events are propagated through a computer model of the BESIII
detector and filtered using the same selection criteria
as for the experimental data.
 The resulting global fit
describes the multidimensional angular distributions very well as
shown in
Figs.~\ref{fitpbarPi} and \ref{fitnbarpi}.
In  a crosscheck the fit was applied 
to the two data sets separately and the obtained values
of the parameters agree within statistical uncertainties 
as shown in Table~\ref{tab:fitpars}. 
 The
details of the fit as well the evaluation of the systematic
uncertainties are discussed in \meth, and 
the contributions to the systematic uncertainty are listed in 
Table~\ref{tab:BGIn:SumAPbarPi}.

\section{Results}
\begin{figure*}[htbp]
\centering
\includegraphics[width=0.49\textwidth]{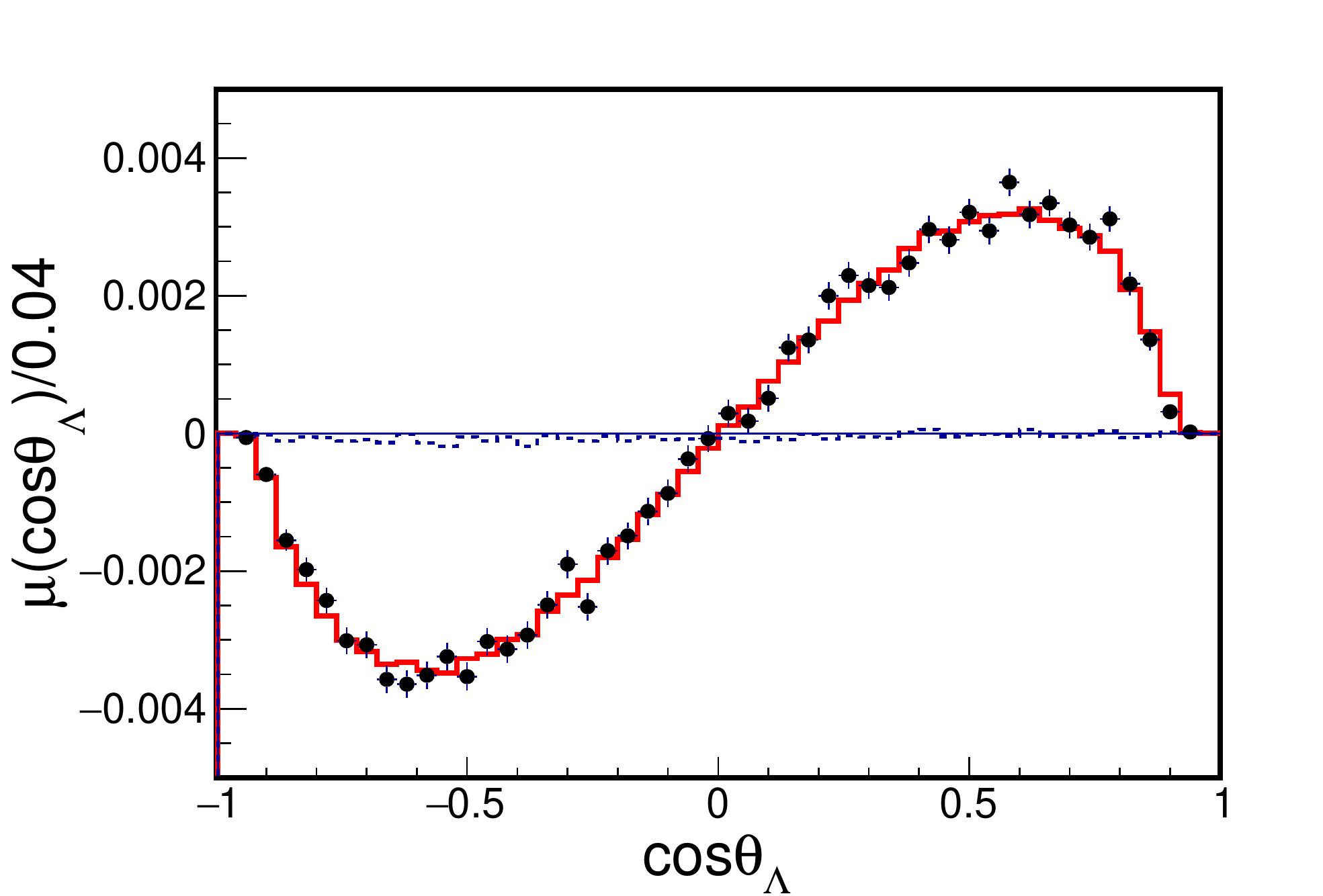}
  \put(-175,115){\large (a)}
\includegraphics[width=0.49\textwidth]{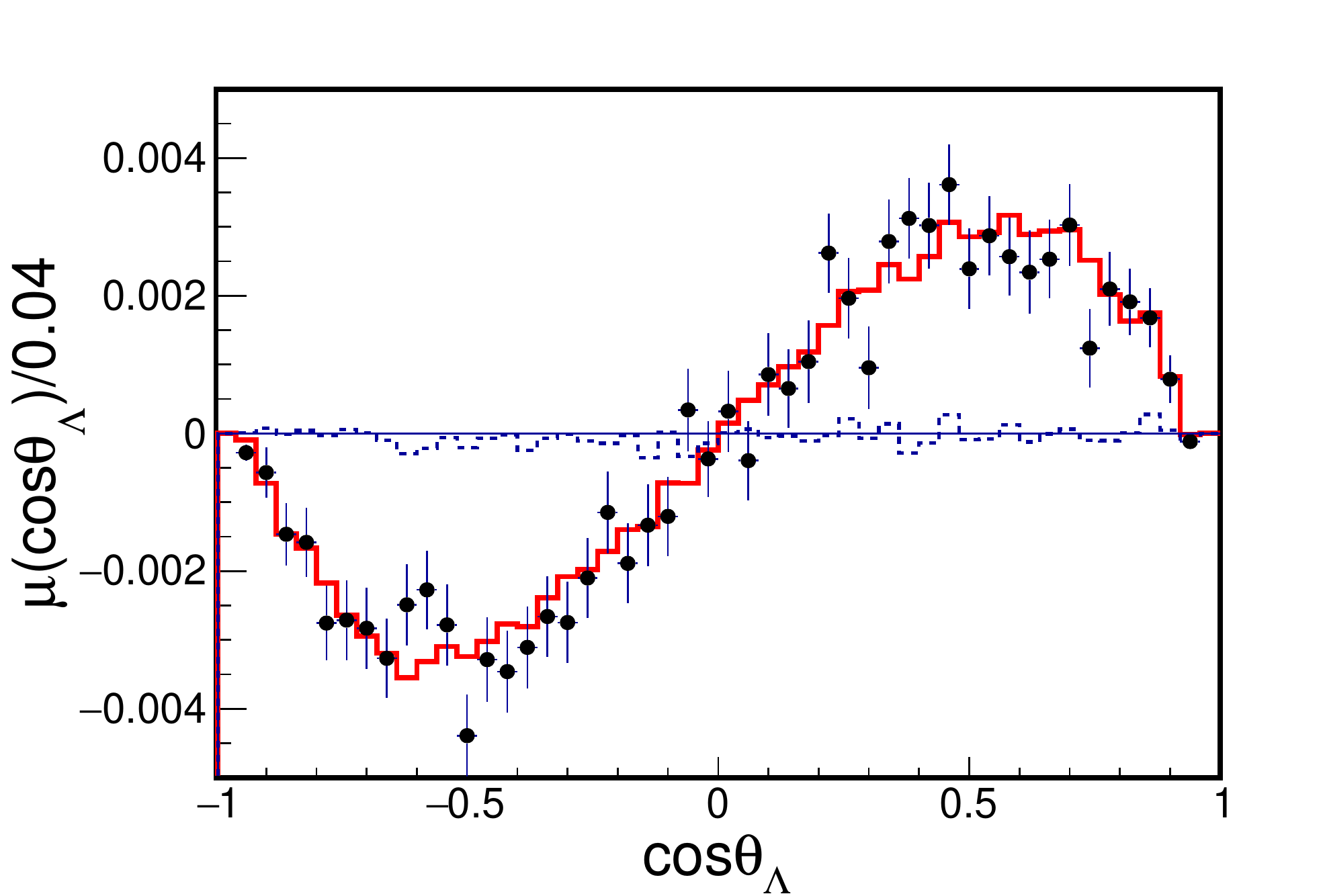}
\put(-175,115){\large (b)}
\caption[]{Moments $\mu(\cos\theta_\Lambda)$ for 
acceptance uncorrected data as a function of $\cos\theta_\Lambda$ for
(a) $p\pi^-\bar{p}\pi^+$ and (b) $p\pi^-\bar{n}\pi^0$ data sets.
The points with error bars are the data, and the solid-line 
histogram is the global 
fit result.
The dashed histogram shows the no polarization scenario (${\cal W}({\boldsymbol{\xi};0,0,0,0})\equiv1$).}
\label{FitPbarPi} 
\end{figure*}
\begin{table}
\centering
\caption[]{Summary of the results:
the $\jpsito\llb$ angular distribution
  parameter $\alpha_\psi$, the phase $\Delta\Phi$, the
  asymmetry parameters for the $\lam\to p\pim$ $(\alpha_-)$, $\bar\Lambda\to\bar
  p\pi^+$ $(\alpha_+)$ and $\bar\Lambda\to \bar n\pi^0$ $(\bar\alpha_0)$ decays, the $CP$ asymmetry
  $A_{CP}$, and the ratio $\bar{\alpha}_{0}/\alpha_{+}$. The first
  uncertainty is statistical, and the second one is systematic.}
  \vspace{0.2cm}
  \renewcommand{\arraystretch}{0.8}
\begin{tabular}{lll}
  \hline  \hline
   Parameters                & \multicolumn{1}{c}{This work} & \multicolumn{1}{c}{Previous results} \\
  \hline
  $\alpha_{\psi}$          &   $\phantom{-}0.461\pm 0.006$  $\pm0.007$ & $\phantom{-}0.469\pm0.027$ \hfill \cite{Ablikim:2017tys}\\
  $\Delta\Phi$ &$\phantom{-}(42.4\pm0.6\pm0.5)^\circ$ & \multicolumn{1}{c}{--} \\\hline

  $\alpha_{-}$               & $\phantom{-}0.750\pm0.009\pm0.004$&$\phantom{-}0.642\pm0.013$\hfill \cite{pdg16}\\

  $\alpha_+$&$-0.758\pm0.010\pm0.007$ & $-0.71\pm0.08$\hfill \cite{pdg16}\\

  $\bar\alpha_0$&$-0.692\pm0.016\pm0.006$ & \multicolumn{1}{c}{--} \\ 

  $A_{CP}$ & $-0.006 \pm 0.012  \pm 0.007 $ &$\phantom{-}0.006\pm0.021$\hfill \cite{pdg16} \\

  $\bar{\alpha}_{0}/\alpha_{+}$& $\phantom{-}0.913\pm 0.028  \pm 0.012$ & \multicolumn{1}{c}{--}\\ 
  \hline  \hline
\end{tabular}
  \vspace{0.3cm}
 \label{tab:BGIn:Sum}
\end{table}
A clear polarization, strongly dependent on the $\Lambda$ direction, $\cos\theta_\Lambda$,
is observed for $\lam$ and $\lbar$. 
In Fig.~\ref{FitPbarPi}, the moment 
$\mu(\cos\theta_\Lambda)=(1/N)\;\sum_i^{N(\theta_\Lambda)}(n_{1,y}^{(i)}-n_{2,y}^{(i)})$,
related to the polarization, is calculated in
50 bins in $\cos\theta_\Lambda$.  $N$ is the total number of events in the data sample and
$N(\theta_\Lambda)$ is the number of events in a $\cos\theta_\Lambda$
bin.  In the limit of CP conservation, $\alpha_-=-\alpha_+$, while an approximate 
isospin symmetry leads to $\alpha_+\approx\bar\alpha_0$ \cite{Olsen:1970vb,Cleland:1972fa}, and the
expected angular dependence is
$\mu(\cos\theta_\Lambda)\sim\sqrt{1-\alpha_\psi^2}\alpha_-\!\sin\Delta\Phi\cos\theta_\Lambda\sin\theta_\Lambda$
for the acceptance corrected data (compare Eq.~\eqref{jointangular}). 
The phase between helicity flip and helicity conserving
transitions
is determined to be
$\Delta\Phi=(42.4\pm0.6\pm0.5)^\circ$, where the first uncertainty 
is statistical and the second systematic.
This large value of the phase 
enables a simultaneous determination of the decay asymmetry parameters for $\lam\ra
p\pim$, $\lbar\ra\pbar\pip$, and $\lbar\ra\nbar\piz$  
as given in Table~\ref{tab:BGIn:Sum}.  The value of
$\alpha_{-}=0.750\pm0.009\pm0.004$ differs by more than five standard
deviations from the
commonly accepted world average value of $\alpha_{-}^\textrm{PDG}=0.642\pm0.013$,
 based on elaborate experiments from
1963-75~\cite{Cronin:1963zb,Overseth:1967zz,Dauber:1969hg,Cleland:1972fa,Astbury:1975hn},
where the daughter proton 
polarization was measured in a secondary scattering process. Our result
means that all published measurements on $\Lambda/\bar\Lambda$
polarization, determined from the product of $\alpha_{-}$ and the
polarization, are $(17\pm3)\%$ too large.  
The obtained value for the  ratio  $\bar{\alpha}_{0}/\alpha_{+}$ is 3$\sigma$ lower than unity, the value expected from the $|\Delta I|
= \frac {1} {2}$ rule for non-leptonic decays of strange
particles~\cite{Crawford:1959zza,Cleland:1972fa}, indicating the importance of radiative corrections \cite{Belavin:1969kd,Olsen:1970vb}.
 The $\afm$
and $\afp$ values determined in this Letter, together with the covariance
matrix, enable a calculation of the $CP$ odd observable
$A_{CP}={(\alpha_-+\alpha_+)}/{(\alpha_--\alpha_+) }
=-0.006\pm0.012\pm0.007$.  This is the most sensitive
test of CP violation for the $\Lambda$ baryon with a
substantially improved precision over previous
measurements~\cite{Barnes:1996si} (Table~\ref{tab:BGIn:Sum})
and using a novel, model independent method.  The
Cabibbo-Kobayashi-Maskawa (CKM) mechanism predicts an $A_{CP}$ value
of $\sim10^{-4}$~\cite{Donoghue:1986hh}, while 
various extensions of the standard
model predict larger value~\cite{Bigi:2017eni}, in an attempt to explain
the observed baryon-antibaryon asymmetry in the universe. 
 This new
method
for CP tests, when applied to the foreseen future larger data samples, can reach
a precision level compatible with theory predictions, which in turn will give a clue 
about baryogenesis.

\acknowledgments 

The BESIII collaboration thanks the staff of BEPCII and the IHEP computing center for their strong support. This work is supported in part by National Key Basic Research Program of China under Contract No. 2015CB856700; National Natural Science Foundation of China (NSFC) under Contracts Nos. 11335008, 11375205, 11425524, 11625523, 11635010, 11735014; the Chinese Academy of Sciences (CAS) Large-Scale Scientific Facility Program; the CAS Center for Excellence in Particle Physics (CCEPP); Joint Large-Scale Scientific Facility Funds of the NSFC and CAS under Contracts Nos. U1532257, U1532258, U1732102, U1732263; CAS Key Research Program of Frontier Sciences under Contracts Nos. QYZDJ-SSW-SLH003, QYZDJ-SSW-SLH040; 100 Talents Program of CAS; the CAS President’s International
Fellowship Initiative; INPAC and Shanghai Key Laboratory for Particle Physics and Cosmology; German Research Foundation DFG under Contracts Nos. Collaborative Research Center CRC 1044, FOR 2359; Istituto Nazionale di Fisica Nucleare, Italy; Koninklijke Nederlandse Akademie van Wetenschappen (KNAW) under Contract No. 530-4CDP03; Ministry of Development of Turkey under Contract No. DPT2006K-120470; National Science and Technology fund; The Swedish Research Council; the Knut and Alice
Wallenberg foundation; U. S. Department of Energy under Contracts Nos. DE-FG02-05ER41374, DE-SC-0010118, DE-SC-0010504, DE-SC-0012069; University of Groningen (RuG) and the Helmholtzzentrum fuer Schwerionenforschung GmbH (GSI), Darmstadt.

\appendix
\section{Methods\label{sec:meth}}
\renewcommand\thefigure{A.\arabic{figure}}    
\renewcommand\thetable{A.\arabic{table}}    
\setcounter{figure}{0} 
\setcounter{table}{0} 
\subsection{Monte Carlo simulation\label{DataMC}}
The optimization of the event selection criteria and the estimation of
backgrounds are based on Monte Carlo (MC) simulations. The {\sc
  geant4}-based simulation software includes the geometry and the
material description of the BESIII spectrometer, the detector response
and the digitization models, as well as the data base of the running
conditions and the detector performance. The production of the
$J/\psi$ resonance is simulated by the MC event generator {\sc
  kkmc}\cite{Jadach:1999vf}; the known decays are generated by {\sc
  besevtgen}\cite{Lange:2001uf,Ping:2008zz} with branching ratios set
to the world average values~\cite{pdg16}, and missing decays are
generated by the {\sc lundcharm}\cite{Chen:2000tv} model with
optimized parameters\cite{Yang:2014vra}.  Signal and background events
are generated using helicity amplitudes. For the signal process,
$J/\psi \to \Lambda \bar{\Lambda}$, the angular distribution of
Eq.~\eqref{jointangular} is used. For the backgrounds, $J/\psi \to
\Sigma^{0} {\bar{\Sigma}}^{0}$, $\Sigma^{+} \bar{\Sigma}^{-}$ and
$\Lambda \bar{\Sigma}^{0} + c.c$ decays, the helicity amplitudes are
taken from Ref.\cite{Zhong:2008rs}, and the angular distribution
parameters are fixed to $-0.24$~\cite{Ablikim:2005cda} for $J/\psi \to
\Sigma^{0} {\bar{\Sigma}}^{0}$ and $J/\psi \to \Sigma^{+}
\bar{\Sigma}^{-}$, and to 0.38~\cite{Ablikim:2012bw} for $J/\psi \to
\Lambda \bar{\Sigma}^{0} + c.c$.
\subsection{Selection Criteria}

\begin{figure*}
\centering
\includegraphics[width=16cm]{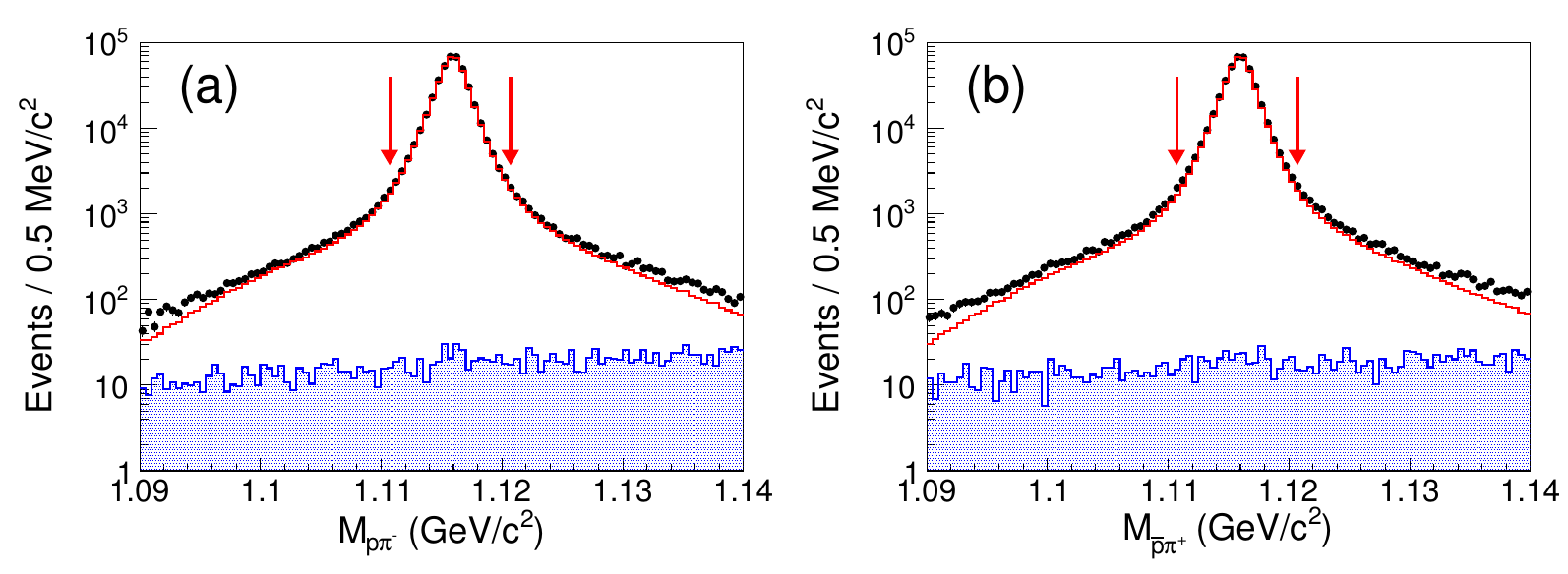}
\caption{Distribution for (a) invariant mass of $p \pi^{-}$ and (b) invariant mass of $\bar p\pi^{+}$. The dots with error bars and the open histogram denote data and signal MC, respectively. The shaded histogram shows the background. The arrows indicate the $\lam$ and $\lbar$ mass windows used to select the signal events.}
\label{LLbarcut} 
\end{figure*}

Charged tracks detected in the Main Drift Chamber (MDC) must satisfy
$|\rm{\cos}\theta|<0.93$, where $\theta$ is the polar angle with
respect to the beam direction.  There are no particle identification
requirements for the tracks.  Showers in the Electromagnetic Calorimeter 
(EMC), not associated with
any charged track, are identified as photon candidates if they fulfill
the following requirements: the deposited energy is required to be larger than 25 MeV
and 50 MeV for clusters reconstructed in the barrel ($|\rm{\cos}\theta|<0.8$) and end cap
($0.86<|\rm{\cos}\theta|<0.92$), respectively.  In order to
suppress electronic noise and showers unrelated to the event, the EMC
time difference from the event start time is required to be within [0,
700] ns. To remove showers originating from charged particles, the angle
between the shower position and charged tracks extrapolated to the
EMC must be greater than 10 degrees.
\paragraph{Selection of \boldmath$\jpsi\to\llb,~\lam\to p\pi^-$, $\lbar\to\bar p\pi^+$}
Events with at least four charged tracks are selected. Fits of the
 $\Lambda$ and $\bar\Lambda$ vertices are performed using all pairs of
 positive and negative charged tracks. There should be at least one
 $\Lambda$$\bar{\Lambda}$ pair in an event. If more than one set of
 $\Lambda$$\bar{\Lambda}$ pairs is found, the one with the smallest value
 of
 $(M_{p\pi^-}-M_{\Lambda})^{2}+(M_{\bar{p}\pi^+}-M_{{\Lambda}})^{2}$,
 where $M_{\Lambda}$ is the nominal 
 $\Lambda(\bar\Lambda)$ mass, is retained for further analysis. A four
 constraint kinematic fit imposing energy-momentum conservation (4C-fit) is
 performed with the $\Lambda\to p \pi^-$ and $\bar{\Lambda}\to\bar{p}\pi^+$ hypothesis, and
 events with $\chi^{2}<60$ are retained.  The invariant masses of
 $p\pi^-$ and $\bar p\pi^+$ are required to be within $|M_{p\pi^-} -
 M_{\Lambda}| <$ 5 $\textrm{MeV}$/$c^{2}$ and $|M_{\bar{p}\pi^+ } -
 M_{{\Lambda}}| <$ 5 $\textrm{MeV}$/$c^{2}$. The $p\pi^-$ and
 $\bar p\pi^+$ invariant mass spectra and the selection windows are
 shown in Fig.~\ref{LLbarcut}.

\begin{figure*}
\centering
\includegraphics[width=16cm]{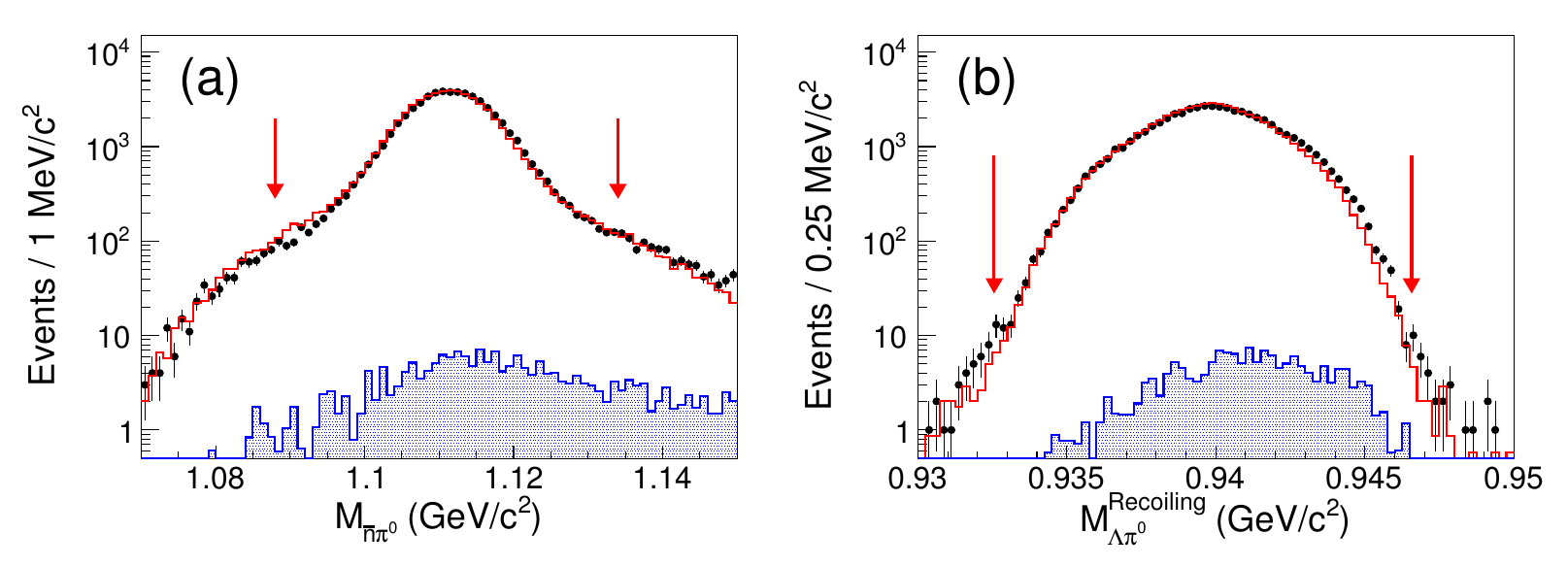}
\caption{Distribution for (a) invariant mass of $\bar{n} \pi^{0}$ and
  (b) recoiling mass of $\Lambda \pi^{0}$. The dots with error bars
  and the open histograms denote data and signal MC, respectively. The
  shaded histograms show the backgrounds. The arrows indicate the
  selection windows.}
\label{NbarPiAlgcut} 
\end{figure*}

\paragraph{Selection of \boldmath$\jpsi\to\llb,~\lam\to p\pi^-,~\lbar\to \bar n \pi^0$}
Events with at least two charged tracks and at least three showers are
selected. Two showers, consistent with being photons, are used to
reconstruct the $\pi^0$ candidates, and the invariant mass of the pair is
required to be in the interval $[0.12, 0.15]$ GeV/$c^2$.  To improve
the momentum resolution, a mass-constrained fit to the $\pi^0$ nominal
mass is applied to the photon pairs, and the resulting energy and
momentum of the $\pi^0$ are used for further analysis.  Candidates
for $\Lambda$ are formed by combining two oppositely charged tracks
into the final states $p \pi^-$. The two daughter tracks are
constrained to originate from a common decay vertex by requiring the
$\chi^2$ of the vertex fit to be less than 100. To identify a $\nbar$
shower, the deposited energy in the EMC is required to be larger than
350 MeV, and the second moment of the cluster is required to be larger
than 20. The moment is defined as $\sum_{i}E_ir_i^2/\sum_i E_i$, where
$E_i$ is the deposited energy in the $i$-th crystal, and $r_i$ is the
radial distance of the crystal $i$ from the cluster center.  To select
the $J/\psi \to \Lambda (p \pi^{-}) \bar\Lambda (\bar{n} \pi^{0})$
candidate events, a one-constraint (1C) kinematic fit is performed,
where the momentum of the anti-neutron is unmeasured.  The selected
events should have the $\chi^{2}_{1C-\bar{n}}$ of the 1C kinematic fit
less than 10, and if there is more than one combination, the one with
the smallest $\chi^{2}_{1C-\bar{n}}$ value is chosen.  To further
suppress background contributions, we require $|M_{p\pi^-} -
M_{\Lambda}| <$ 5 $\textrm{MeV}$/$c^{2}$, where $M_{\Lambda}$ is the $
\Lambda$ nominal mass.  Figure~\ref{NbarPiAlgcut} shows the invariant
mass ($M_{\bar{n} \pi^{0}}$) of the $\nbar\piz$ pair and the mass
$M^\textrm{Recoiling}_{\Lambda \pi^{0}}$ recoiling against the  $\Lambda \pi^{0}$, where
$M_{\bar{n} \pi^{0}} = \sqrt{(E_{\bar{n}}+E_{\pi^0})^2 -
(\vec{P}_{\bar{n}} + \vec{P}_{\pi^0})^2}$, $\vec{P}_{\bar{n}}=
-(\vec{P}_{\Lambda}+\vec{P}_{\pi^0})$ is in the rest frame of J$/\psi$,
and $E_{\bar{n}} = \sqrt{|\vec{P}_{\bar{n}}|^2 + M_{{n}}^2}$ (with
$M_{{n}}$ the nominal neutron mass).  The signal regions are
defined as $|M_{\bar{n} \pi^{0}} - M_{{\Lambda}}| <$ 23
$\textrm{MeV}$/$c^{2}$ and $|M^\textrm{Recoiling}_{\Lambda \pi^{0}} -
M_{{n}}| <$ 7 $\textrm{MeV}$/$c^{2}$ as shown in
Fig.~\ref{NbarPiAlgcut}.

\paragraph{Background analysis\label{bkg2}}
The potential backgrounds are studied using the inclusive MC sample
for $\jpsi$ decays.  After applying the same selection criteria as for
the signal, the main backgrounds for the $\lbar\to\bar p\pi^+$ final
state are from $J/\psi \to \gamma \llb$, $\Lambda
\bar{\Sigma}^{0} + c.c.$, $\Sigma^{0} \bar{\Sigma}^{0}$, $\Delta^{++}
\bar{p} \pi^{-} + c.c.$, $\Delta^{++} \bar\Delta^{--}$, and $p \pi^{-}
\bar{p} \pi^{+}$ decays. Decays of $J/\psi \to \Lambda
\bar{\Sigma}^{0} + c.c.$ and $\Sigma^{0} \bar{\Sigma}^{0}$ are
generated using the helicity amplitudes and include subsequent
$\Lambda$ and $\bar\Lambda$ decays. The remaining decay modes are
generated according to the phase space model, and the contribution is
shown in Fig.~\ref{LLbarcut}. For the $\lbar\to \bar n \pi^0$ final
state, the dominant background processes are from the decay modes
$J/\psi \to \gamma \llb$, $\Lambda \bar{\Sigma}^{0} + c.c.$,
$\Sigma^{0}(\gam\Lambda) \bar{\Sigma}^{0}(\gam\lbar)$,
$\Sigma^{+}(p\piz) \bar{\Sigma}^{-}(\nbar\pim)$,
$\Lambda(p\pi^-)\bar\Lambda(\bar p\pi^+)$. Exclusive MC samples for
these backgrounds are generated and used to estimate the background
contamination shown in Fig.~\ref{NbarPiAlgcut}.

\subsection{The global fit\label{fitmethod1}}
\begin{figure*}
\centering
\includegraphics[width=17cm]{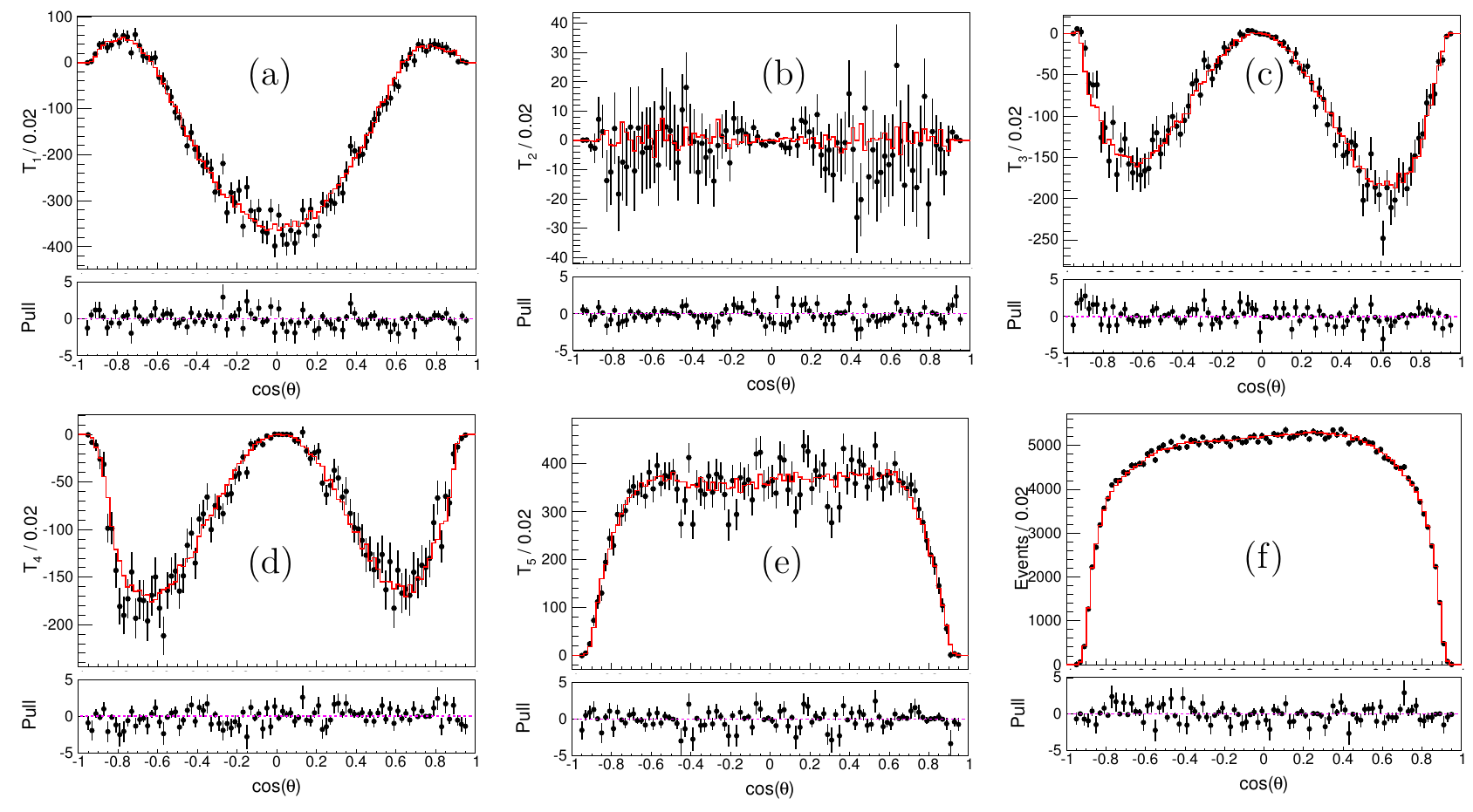}
\caption{Distributions of $T_i~(i=1,...,5)$ functions in terms of
$\cos\theta_\lam$, and the $\lam$ angular distribution for the decays
$\jpsi\to\llb,~\lam\to p\pi^-$ and $\lbar\to \bar p\pi^+$. The dots
with error bars are the data, and the histograms are the total fit
results. The pull distribution is shown at the bottom of each
histogram.}
\label{fitpbarPi} 
\end{figure*}

\begin{figure*}
\centering
\includegraphics[width=17cm]{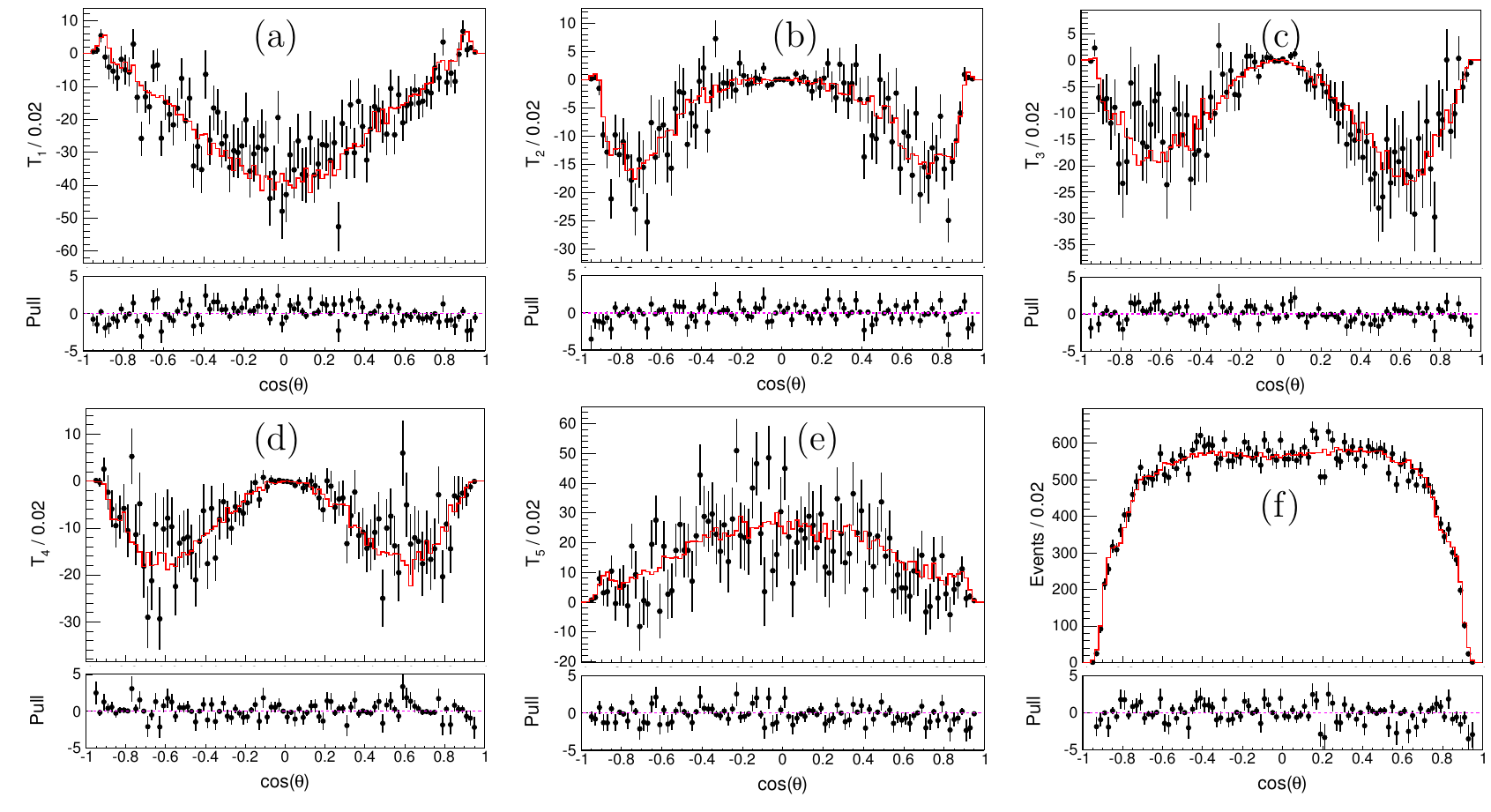}
\caption{Distributions of $T_i~(i=1,...,5)$ functions in terms of
  $\cos\theta_\lam$, and the $\lam$ angular distribution for decays
  $\jpsi\to\llb,~\lam\to p\pi^-$ and $\lbar\to \nbar\pi^0$. The dots
  with error bars are the data, and the histograms are the total fit
  results. The pull distribution is shown at the bottom of each plot.}
\label{fitnbarpi} 
\end{figure*}

Based on the joint angular distribution shown in Eq.~\eqref{jointangular}, a simultaneous fit is performed to the two data sets according to the decay modes:\\
\begin{eqnarray}
&\textrm{I:}&~  \jpsi\to\lam\lbar,~\lam\to p\pi^-  \textrm{~and~} \lbar\to\bar p\pi^+,\nonumber\\
&\textrm{II:}&~ \jpsi\to\lam\lbar,~\lam\to p\pi^- \textrm{~and~} \lbar\to \bar n\pi^0.\nonumber
\end{eqnarray}
There are three common parameters, $\alpha_\psi$, $\Delta\Phi$ and $\afm$, and two separate parameters $\afp$ and $\afz$ for the $\lbar$ decays to $\pbar \pi^+$ and $\nbar\pi^0$, respectively.
For data set I, the joint likelihood function is defined as\cite{Zhong:2008rs}:
\begin{equation}
      {\cal L^\textrm{I}} = \prod^{N^\textrm{I}}_{i=1}{\cal P}({\xi}^{(i)}_\textrm{I}) = ({\cal{C}^\textrm{I}})^{N^\textrm{I}}\prod^{N^\textrm{I}}_{i=1}\mathcal {W}({\xi}^{(i)}_\textrm{I};\alpha_\psi,\Delta\Phi,\afm,\afp),
\end{equation}
where ${\cal P}({\xi}^{(i)}_\textrm{I})$ is the probability density
function evaluated for event
$i$, and ${\mathcal W}({{\xi}^{(i)}_\textrm{I};\alpha_\psi,\Delta\Phi,\afm,\afp})$ is calculated with
Eq.~\eqref{jointangular} for event $i$. The normalization factor,
$({\cal
C}^\textrm{I})^{-1}=\frac{1}{N_\textrm{MC}}\sum_{j=1}^{N_\textrm{MC}}{\mathcal
W}({\xi}^{(j)}_\textrm{I};\alpha_\psi,\Delta\Phi,\afm,\afp)$, is estimated with the accepted
$N_\textrm{MC}$ events, which are generated with the phase space
model, undergo detector simulation, and are selected with the same event
criteria as for data.  The definition of the likelihood function for
data set II, $\cal{L}^\textrm{II}$, is the same except for its
calculation with different parameters and data set. To determine the
parameters, we use the package MINUIT from the CERN library
\cite{James:1975dr} to minimize the function defined as:
\begin{equation}\label{subtractBG}
S= -\ln {\cal L}^\textrm{I}_{\textrm{data}}-\ln {\cal L}^\textrm{II}_{\textrm{data}}+\ln \cal{L}^\textrm{I}_{\textrm{bg.}}+\ln \cal{L}^\textrm{II}_{\textrm{bg.}},
\end{equation}
where $\ln{\cal L}^\textrm{I(II)}_{\textrm{data}}$ and $\ln \cal{L}^\textrm{I(II)}_{\textrm{bg.}}$ are the likelihood functions for the two data sets and the background events, respectively. The results of the separate fits  for the two data sets 
are given in Table~\ref{tab:fitpars}.
We compare the fit with the data using moments $T_i~(i=1,...,5)$ directly
related 
to the terms in Eq.~\eqref{jointangular}.  The moments 
 are explicitly given by
\begin{eqnarray}
T_1&=& \sum_i^{N(\theta_\Lambda)}\left(\sin ^2\theta_\Lambda n_{1,x}^{(i)} n_{2,x}^{(i)}+\cos^2\theta_\Lambda  n_{1,z}^{(i)}n_{2,z}^{(i)}\right),\nonumber\\
T_2&=& -\sum_i^{N(\theta_\Lambda)}\sin  \theta_\Lambda\cos  \theta_\Lambda( n_{1,x}^{(i)}n_{2,z}^{(i)}
+  n_{1,z}^{(i)}n_{2,x}^{(i)})\nonumber, \\
T_3&=& -\sum_i^{N(\theta_\Lambda)}\sin \theta_\Lambda\cos\theta_\Lambda n_{1,y}^{(i)},\nonumber\\
T_4&=& -\sum_i^{N(\theta_\Lambda)}\sin \theta_\Lambda\cos\theta_\Lambda  n_{2,y}^{(i)},\nonumber\\
T_5&=&\sum_i^{N(\theta_\Lambda)}\left(n_{1,z}^{(i)}n_{2,z}^{(i)}- \sin ^2\theta_\Lambda n_{1,y}^{(i)}n_{2,y}^{(i)}\right).\nonumber
\end{eqnarray}
Figs.~\ref{fitpbarPi} and \ref{fitnbarpi} show the fit results to data
sets I and II, respectively.  The distributions of $T_i~(i=1,...,5)$
functions in terms of $\cos\theta_\Lambda$, and the $\lam$ angular
distribution are shown in histograms (a)-(f), together with the
corresponding pull distributions. The asymmetric distributions of
$T_3$ and $T_4$ indicate that significant transverse polarization of
$\Lambda$ and $\lbar$ hyperons is observed.  The simultaneous fit
results for $\alpha_{\psi},~\alpha_-,~\alpha_+,~\Delta\Phi$ and
$\bar\alpha_0$ parameters are given in Table \ref{tab:fitpars}. Based
on these parameters, the observables $\bar\alpha_0/\alpha_+$ and
$A_{CP}=({\alpha_-+\alpha_+})/({ \alpha_--\alpha_+})$ are calculated,
and their statistical uncertainties are evaluated taking into account
correlation coefficients $\rho(\alpha_+,\alpha_0)=0.42$ and
$\rho(\alpha_+,\alpha_-)=0.82$, respectively. As a cross check,
separate fits to data sets I and II are performed, and results are
consistent with the simultaneous fit within statistical uncertainties,
as given in Table \ref{tab:fitpars}.

\begin{table}
\centering
\caption{Simultaneous fit results for the angular distribution parameter, $\alpha_{\psi}$,  $\Delta\Phi$
 and the asymmetry parameters $\alpha_-$ for decays $\lam\to p\pim,$ $\alpha_+$ for $~\bar\Lambda\to\bar p\pi^+$ and $\alpha_0$ for $\bar\Lambda\to \bar n\pi^0$, and compared to the separate fit results to data I and II samples. The uncertainties are statistical only.}
%
\vspace{0.2cm}
\renewcommand{\arraystretch}{0.8}
\begin{tabular}{cccc}
  \hline  \hline
   Pars.         & Simultaneous fit & Fit to I & Fit to II  \\
  \hline
  $\alpha_{\psi}$   & $0.461\pm 0.006$ & $0.459\pm0.006$ &$0.473\pm0.019$\\

  $\alpha_{-}$        & $0.750\pm0.009$  &$0.749\pm0.010$  &$0.756\pm0.031$\\

  $\alpha_+$          &$-0.758\pm0.010$  & $-0.759\pm0.010$ &......\\

  $\bar\alpha_0$      &$-0.692\pm0.016$  & ......           &$-0.684\pm0.028$\\

  $\Delta\Phi$ $(^\circ)$     &$42.4\pm0.6$   & $42.3\pm0.6$ & $43.4\pm2.1 $  \\\hline

\end{tabular}
\vspace{0.3cm}
\label{tab:fitpars}
\end{table}


\subsection{Systematic uncertainty}\label{sec:syserr}

The systematic uncertainties can be divided into two categories. The first category is from the event selection, including the uncertainties on  MDC tracking efficiency, the kinematic fit, $\pi^0$ and $\bar{n}$ efficiencies, $\Lambda$ and $\bar \Lambda$ reconstruction, background estimations, and the $\Lambda$,  $\bar \Lambda$ and $M^\textrm{Recoiling}_{\Lambda \pi^{0}}$ mass window requirements. The second category includes uncertainties associated with the fit procedure.

\begin{enumerate}

\item The uncertainty due to the efficiency of the charged particles
  tracking has been investigated with $J/\psi \to \Lambda
  \bar{\Lambda} \to p \pi^{-} \bar{p} \pi^{+}$ control 
samples\cite{Ablikim:2014cea}, taking into consideration the correlation between the
  magnitude of charged particle momentum and its polar angle
  acceptances. Corrections are made based on the 2-dimensional
  distribution of momentum versus polar angle. The difference between
  the fit results with and without the tracking correction is taken as
  a systematic uncertainty.

\item The uncertainty due to the $\piz$ reconstruction is estimated
  from the difference between data and MC simulation using a $J/\psi
  \to \pi^{+} \pi^{-} \pi^{0}$ control sample. The uncertainty due to
  $\bar{n}$ shower requirement is estimated with a $J/\psi \to p
  \pi^{-} \bar{n}$ control sample, and the correction factors between
  data and MC simulations are determined.  The differences in the fit
  results with and without corrections to the efficiencies of the
  $\pi^0$ and $\bar n$ reconstructions are taken as systematic
  uncertainties.

\item The systematic uncertainties for the determination of the
  physics parameters in the fits due to the $\Lambda$ and $\lbar$
  vertex reconstructions are found to be negligible.

\item The systematic uncertainties due to kinematic fits are
  determined by making corrections to the track parameters in the MC
  simulations to better match the data. The corrections are done with the
  5-dimensional distributions over the $\theta_\Lambda$, $\hat{\bf
    n}_1$, $\hat{\bf n}_2$ variables, where the $\hat{\bf n}_1$ and
  $\hat{\bf n}_2$ are expressed using spherical coordinates. The fit
  to data with the corrected MC sample yields $\alpha_{\psi}$=0.462
  $\pm$ 0.006, $\alpha_{-}=0.749\pm0.009,~\alpha_{+} =- 0.752 \pm
  0.009$, and $\bar\alpha_0=-0.688\pm0.017$. The differences between
  the fit with corrections and the nominal fit are considered as the
  systematic uncertainties. For $\alpha_{\psi}$, the difference
  between the fit results with and without this correction is
  negligible.

\item The uncertainty due to the fit method is estimated with MC simulations. The differences between the input and output values on the physical parameters  are taken as the systematic uncertainties.

\item The systematic uncertainty caused by the background estimation is studied by fitting the data with and without considering background subtraction. The differences on the parameters are taken as the systematic uncertainties. The contamination rate of background events in this analysis is less than 0.1\% according to the full MC simulations, and the uncertainty due to background estimation is negligible.
\end{enumerate}
The total systematic uncertainty for the parameters is obtained by
summing the individual systematic uncertainties in quadrature, and is
summarized in Table~\ref{tab:BGIn:SumAPbarPi}.

\begin{table}
\centering
\caption{ Relative systematic uncertainties (\%) for the measurements of parameters $\ajpsi,~\alpha_{-},~\alpha_{+},~\bar\alpha_0$ and $\Delta\Phi$. Uncertainties due to the $\Lambda/\bar\Lambda$ vertex and backgrounds are ignored.}
%
  \vspace{0.2cm}
  \renewcommand{\arraystretch}{0.7}
\begin{tabular}{l l l l l l}
  \hline  \hline
  Source                           & $\ajpsi$    &  $\alpha_{-}$ & $\alpha_{+}$& $\bar\alpha_0$ &$\Delta\Phi$\\
  \hline
  Tracking,$\pi^0$, $\nbar$                               & 1.5  & 0.1 & 0.3 &0.6 &1.1\\

  Kinematic fit                  & 0.2    & 0.1 &0.8&0.6& 0.0\\

  Fit method                                    & 0.0    & 0.5 &0.4 &0.4 &0.1\\
  \hline
  Total                             & 1.5   & 0.5& 0.9 & 0.8 &1.1 \\
  \hline  \hline
\end{tabular}
  \vspace{0.3cm}
 \label{tab:BGIn:SumAPbarPi}
\end{table}

\bibliography{llpap} 
\end{document}